\documentclass[useAMS,usenatbib]{mn2e}
\usepackage{graphicx,amssymb}
\usepackage{mnras_macros}
\title[The origin of dark satellite galaxies]
{The origin of failed subhaloes and the common mass scale of the Milky Way satellite galaxies}
\author[T. Okamoto and C. S. Frenk]{Takashi Okamoto$^{1, 2}$\thanks{E-mail:
tokamoto@ccs.tsukuba.ac.jp}, Carlos S. Frenk$^{2}$\\
$^{1}$ Center for Computational Sciences, University of Tsukuba, 1-1-1 
 Tennodai, Tsukuba 305-8577 Ibaraki, Japan\\  
$^{2}$Institute for Computational Cosmology, Department of Physics,
Durham University, South Road, Durham, DH1 3LE}
\begin{document}

\date{Accepted . Received ; in original form }

\pagerange{\pageref{firstpage}--\pageref{lastpage}} \pubyear{2008}

\maketitle

\label{firstpage}

\begin{abstract}
We study the formation histories and present-day structure of
satellite galaxies formed in a high resolution hydrodynamic simulation
of a Milky Way-like galaxy. The simulated satellites span nearly 4
orders of magnitude in luminosity but have a very similar mass within
their inner 600 pc, $\sim 3 \times 10^7 \ M_\odot$, with very little
scatter. This result is in agreement with the recent measurements for
dwarf spheroidal galaxies (dSphs) in the Milky Way by Strigari et al. In our
simulations a preferred mass scale arises naturally from the effects
of the early reionisation of gas. These impose a sharp threshold of
$\sim 12$ km s$^{-1}$ on the circular velocity of haloes which can cool
gas and make stars. At the present day, subhaloes that host satellites
as luminous as the classical Milky Way dwarfs ($L_{\rm V}\ge 2.6
\times 10^5 \ L_\odot$),  have typically grown to have circular
velocities of $\gtrsim 20 \ {\rm km} \ {\rm s}^{-1}$. There are,
however, subhaloes with similar circular velocities today which were,
nevertheless, below threshold at reionisation and thus remain
dark. Star formation in above-threshold haloes is truncated when the
halo is accreted into the main galaxy progenitor. Thus, many
properties of today's dwarf satellites such as their luminosity and star
formation rate are set by their accretion time. 

\end{abstract}

\begin{keywords}
methods: numerical --  galaxies: evolution -- galaxies: formation  --
cosmology: theory -- galaxies: dwarf. 
\end{keywords}

\section{Introduction}

The mass density of the Universe is dominated by exotic dark matter
whose identity remains unknown. Important clues could be hidden in the
structure of the dwarf satellite galaxies of the Milky Way which are
the most dark matter dominated systems known in the Universe. It has
recently been found that these galaxies all seem to
have approximately the same central density (within 300 or 600~pc),
irrespective of their luminosity \citep{mat98, gil07, str07,
str08}. There thus appears to be a special scale or threshold involved
in the formation of dwarf satellites. A threshold central density
could arise naturally if there is a cutoff in the primordial power
spectrum of density fluctuations produced by free-streaming of warm
dark matter particles in the early universe \citep{bond83}. However,
if the dark matter is cold, then there are no primordial processes
that operate on the relevant scales. In this case the origin of the
threshold must be a late-time astrophysical process that inhibits star
formation in small dark matter haloes, such as the reionisation of gas
at early times
\citep*[e.g.][]{efs92, ogt08}. 

By suppressing star formation in small cold dark matter haloes, early
reionisation would also solve the `missing satellite' problem
\citep{kly99, moo99}, as demonstrated using semi-analytic models of
galaxy formation \citep{bkw00, ben02, som02, kravtsov04}.  
Recently, \cite{maccio09a} and \cite{li09} have confirmed, 
also with semi-analytic techniques, that this process sets the
common mass scale seen in the inner 300 (or 600) pc of dwarf
satellite haloes.  One problem with the work of \cite{li09} and
\cite{maccio09a}, however, is that they modelled reionisation using
the filtering mass approach of \cite{gne00} which has been shown to
overestimate the importance of this effect \citep{hoe06, ogt08}.

Since the reionisation of gas at early times and its effects on star
formation are complex processes, full hydrodynamic simulations are
required to determine how they influence the structure and evolution
of dwarf galaxies. In this Letter, we analyse a high resolution
hydrodynamic simulation that follows the formation of a Milky Way-type
galaxy and is well suited to a detailed study of the formation of
dwarf satellite galaxies. This simulation has been shown by 
\citet[][Fig.~10]{ofjt09}
to give an excellent match to the observed number of faint
satellites in the Milky Way down to the resolution limit of the
simulation ($V$-band magnitude $M_V=-7$), although it predicts
somewhat too few very bright ($M_V<-18$) satellites. 
Here, we focus on the physical processes that determine
whether or not a subhalo is able to form a galaxy and on the inner
structure of the resulting object.

\section{The simulations}

We study the satellite population that formed in a high-resolution
resimulation, using full baryonic physics, of `Aquarius-D', one of the
6 cold dark matter haloes simulated as part of the `Aquarius' project
by \citet{aquarius}. Our simulation includes all the gas physics
thought to be relevant for galaxy formation: a time-evolving
photoionising background \citep{hm01}, metallicity-dependent gas
cooling and photoheating \citep{wss09}, SN feedback
\citep{onb08}, and chemical evolution due to SNe~II, SNe~Ia, and AGB stars
\citep{oka05}. Details of this simulation may be found in
\citet[][their halo `Aq-D-HR']{ofjt09}. 

For our analysis two processes are particularly important:
reionisation and galactic winds. The former is treated  as
described in \citet{ogt08}. For the latter, we use the model introduced by
\citet{ofjt09}. In brief, when a gas particle
receives an amount of energy $\Delta E$ during a time-step $\Delta
t$, this particle is added to a wind with probability, $p_{\rm w} =
\Delta E/(\frac{1}{2} m_{\rm gas} v_{\rm w}^2)$, where $v_{\rm w}$ is
the initial wind speed. This is given as $v_{\rm w} = 5 \sigma$, where
$\sigma$ is the one-dimensional velocity dispersion of dark matter
particles around the gas particle. This expression implies that the
wind mass generated by an SN is proportional to $\sigma^{-2}$, i.e. per
unit of star formation, less massive galaxies blow more (but slower) 
winds. This model reproduces both the observed luminosity function of
the Milky Way satellites and the luminosity-metallicity relation. 

The mass of dark matter and gas particles in the simulation are
$m_{\rm DM} \simeq 1.5 \times 10^5 \ h^{-1} M_\odot$ and $m_{\rm gas}
\simeq 3.5 \times 10^4 \ h^{-1} M_\odot$, respectively.  The
gravitational softening length for the dark matter, gas, and star
particles is kept fixed in comoving coordinates for $z > 3$;
thereafter it is frozen in physical units, $\epsilon = 0.175 \
h^{-1}$kpc. To test for numerical convergence, we also ran a low
resolution version of the simulation with $m_{\rm DM} \simeq 1.9
\times 10^6 \ h^{-1} M_\odot$, $m_{\rm gas}
\simeq 4.1 \times 10^5 \ h^{-1} M_\odot$, and $\epsilon = 0.425 \
h^{-1}$kpc.

In order to identify subhaloes and satellites, we use the SUBFIND
algorithm \citep{spr01}. We define subhaloes as systems that consist
of at least 32 particles, and satellites as subhaloes that contain at
least 10 star particles, unless otherwise stated. Satellite
luminosities are computed using the population synthesis code
P$\acute{\rm E}$GASE2 \citep{pegase}.

\section{Results} 

\begin{figure}
\begin{center}
\includegraphics[width=8.0cm]{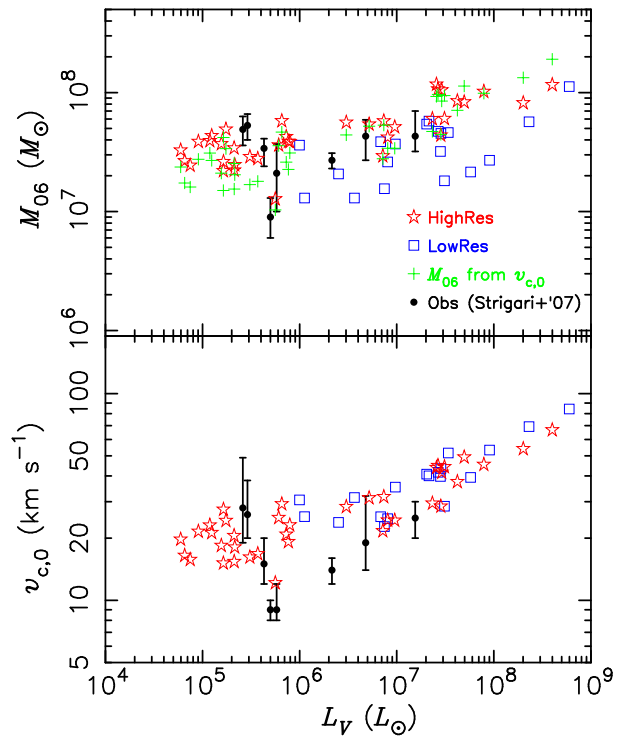}
\end{center}
\caption{ 
  Upper panel: mass within the inner 600 pc of satellite galaxies as a
  function of $V$-band luminosity. The stars and squares show the mass
  obtained directly from the high- and low-resolution simulations,
  respectively. The points with error bars show measurements for the
  classical dSphs of the Milky Way
  \citep{str07}. The values of $M_{06}$ derived from the empirical
  relation between $M_{06}$ and $v_{\rm c, 0}$ 
  are indicated by the plus signs. Lower panel: as the upper panel, but for the
  maximum of the circular velocity profile of satellites, $v_{\rm c, 0}$. }
\label{strigari}
\end{figure}

The main result of our study is presented in Fig.~\ref{strigari}
where, in the upper panel, we plot the mass within the inner 600 pc,
$M_{06}$, of the 38 satellites resolved in the
simulation, as a function of the $V$-band luminosity. The mass varies
by less than a factor of 5 over 4 orders of magnitude in luminosity,
reproducing very well the overall behaviour of the
\cite{str07} data for classical dSphs, 
shown by the black dots. The lower panel shows the maximum of the
circular velocity curve, $v_{\rm c, 0} = v_{\rm c}(r_{\rm max})$,
which also agrees well with the data. The latter agreement is perhaps
not too surprising given that in order to derive values of $v_{{\rm
c}, 0}$ for the Milky Way satellites, \cite{str07} adopted a relation
between $v_{{\rm c}, 0}$ and $r_{\rm max}$ derived from the
$\Lambda$CDM simulation of
\cite{Diemand07b}.
 
Before examining the physical processes responsible for the  
result of Fig.~\ref{strigari}, we discuss the reliability of our
numerical data. Both panels of this figure include results from our
low resolution simulation. For $v_{\rm c, 0}$ these agree very well
with those from the high resolution simulation, indicating that this
quantity has converged. For $M_{06}$ on the other hand, the results
from the low resolution simulation are about a factor of 2 lower than
for the high resolution simulation. The explanation of this apparent
paradox is simply that, $v_{\rm c, 0}$ 
is numerically more robust than $M_{06}$ \citep{aquarius}. 
To test whether or
not $M_{06}$ has converged in our high resolution simulation, we make
use of the empirical relation between $M_{06}$ and $v_{\rm c, 0}$
derived from the highest resolution Aquarius 
simulation\footnote{$\log[v_{\rm c, 0}/({\rm km~s}^{-1})] = 
0.58\log[M_{06}/(h^{-1}M_\odot)] - 3.092$} 
(which has about 100 times better resolution than our high resolution
hydrodynamic simulation; \citet{aquarius}). As shown in
Fig.~\ref{strigari}, the values of $M_{06}$ derived from this relation
agree well with the directly measured values, showing that the high
resolution simulation gives converged and reliable values of $M_{06}$
for all the satellites resolved in the simulation.

\begin{figure}
\begin{center}
\includegraphics[width=8.0cm]{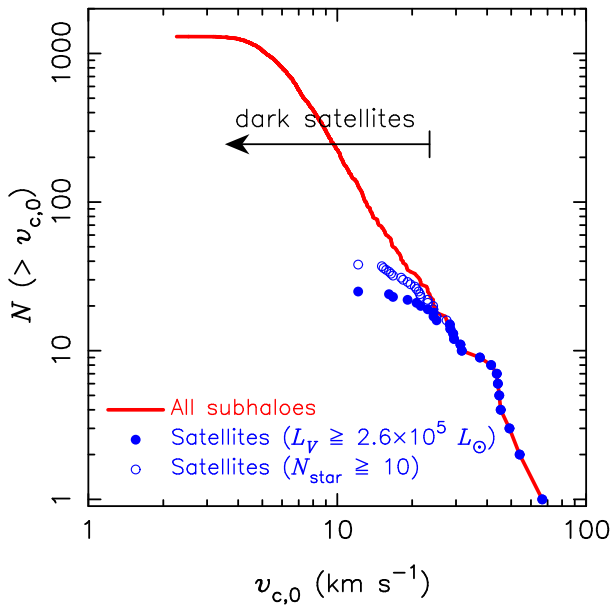}
\end{center}
\caption{The cumulative distribution of maximum circular velocity for 
  all subhaloes and visible satellites. The solid line gives
  the number of dark matter subhaloes in the halo and the solid
  circles show the number of simulated satellites that are brighter
  than Draco. The distribution for the full satellite sample ($N_{\rm
  star} \ge 10$) is shown with open circles. The black arrow
  indicates the value of the circular velocity (23.5 km s$^{-1}$)
  below which failed subhaloes, i.e. those that contain no star
  particles in the simulation, exist. }
\label{vf}
\end{figure}

The lower panel of Fig.~\ref{strigari} provides an important clue to 
the reasons behind the preferred mass scale picked out in the top
panel. Over the range spanned by the observed dSphs 
($L_{\rm V} \lesssim 2 \times10^7 L_\odot$), the velocity
$v_{\rm c, 0}$ is essentially constant, at $\sim 20$ km s$^{-1}$, and
then it rises slowly with luminosity at brighter magnitudes. Thus,
there appears to be a threshold circular velocity required for a dwarf
satellite to form. This threshold is evident in Fig.~\ref{vf} which
compares the circular velocity distribution of satellites to that of
the entire subhalo population. All subhaloes with $v_{\rm c, 0}
\gtrsim 25$ km s$^{-1}$ succeed in making a satellite, but below this
value, the fraction of successful subhaloes drops rapidly. This 
inability of small subhaloes to form a visible galaxy is the reason why
there is no `satellite problem' in the CDM model: only a small
fraction of the large subhalo population are lit up by galaxy
formation.

%

\begin{figure}
\begin{center}
\includegraphics[width=8.0cm]{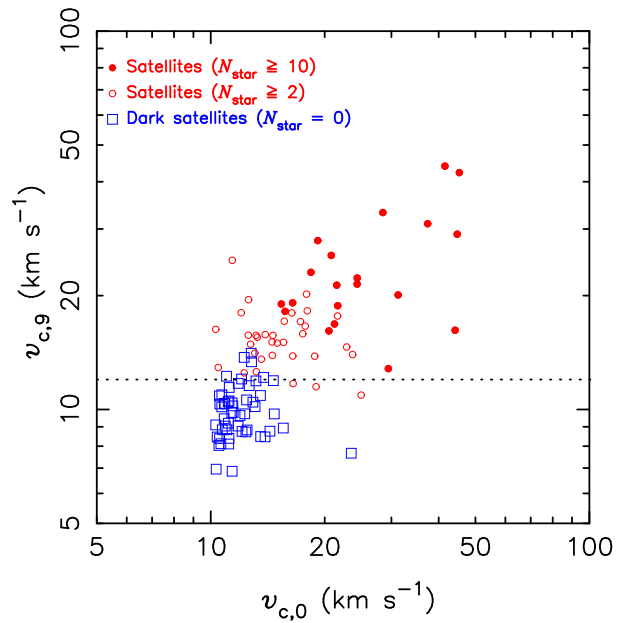}
\end{center}
\caption{ The maximum circular velocity, $v_{\rm c, 9}$, at the epoch of 
  reionisation, $z_{\rm re} = 9$, of the progenitors of present-day
  subhaloes, plotted against their maximum circular velocity today,
  $v_{\rm c, 0}$. Satellite galaxies containing at least ten star
  particles at $z = 0$ are represented by the solid circles while dark
  satellites that do not contain any star particles at $z = 0$ are
  shown by squares. We also show subhaloes containing at least two
  star particles at $z = 0$ by open circles. There is a threshold
  value of $v_{\rm c, 9}$ that separates successful from failed
  subhaloes. For reference, we represent the threshold by the dotted
  line at $v_{\rm c, 9} = 12$ km s$^{-1}$. }
\label{vc9}
\end{figure}

Note, however, that there is no hard cutoff in $v_{\rm c, 0}$ in
Fig.~\ref{vf}. Below $v_{\rm c, 0} = 23.5$ km s$^{-1}$, there are
subhaloes that succeed in making satellites as luminous as the
classical dSphs ($L_{\rm V} \ge 2.6 \times 10^5
\ L_\odot$), but also {\it failed} subhaloes that do not contain any
stars. The difference between haloes of similar, relatively large
circular velocities which do and do not form a visible galaxies is
apparent in Fig.~\ref{vc9}. Here we compare the {\em present day}
maximum circular velocity of a subhalo, $v_{\rm c, 0}$, with the
maximum circular velocity of its main progenitor, $v_{\rm c, 9}$, at
the {\em epoch of reionisation}, which in our simulations happens at
$z_{\rm re} = 9$. This figure shows a sharp transition in $v_{\rm c,
9}$ between subhaloes whose descendants go on to make visible galaxies
and those whose descendants do not. Surprisingly, this threshold
applies not only to luminous satellites, but also to subhaloes that
contain only two star particles. The transition occurs at $v_{\rm c,
9} \simeq 12$ km s$^{-1}$ and is due to the effects of reionisation:
the increase in thermal pressure which follows reionisation evaporates
the gas from subthreshold haloes
\citep[e.g.][]{ree86, ogt08}.

Haloes with maximum circular velocity above the transition critical
velocity, $v_{\rm c, 9} > 12$ km s$^{-1}$, are observed as dwarf
satellites today. Their circular velocities today peak at around
$v_{\rm c, 0} \sim 20$ km s$^{-1}$, as may be seen in Fig.~\ref{vc9}.
This is the origin of the common velocity scale seen in the lower
panel of Fig.~\ref{strigari}. This characteristic velocity translates
directly into the characteristic inner halo mass measured by observers
in today's satellites \citep{gil07, str07, str08}.

\begin{figure}
\begin{center}
\includegraphics[width=8cm]{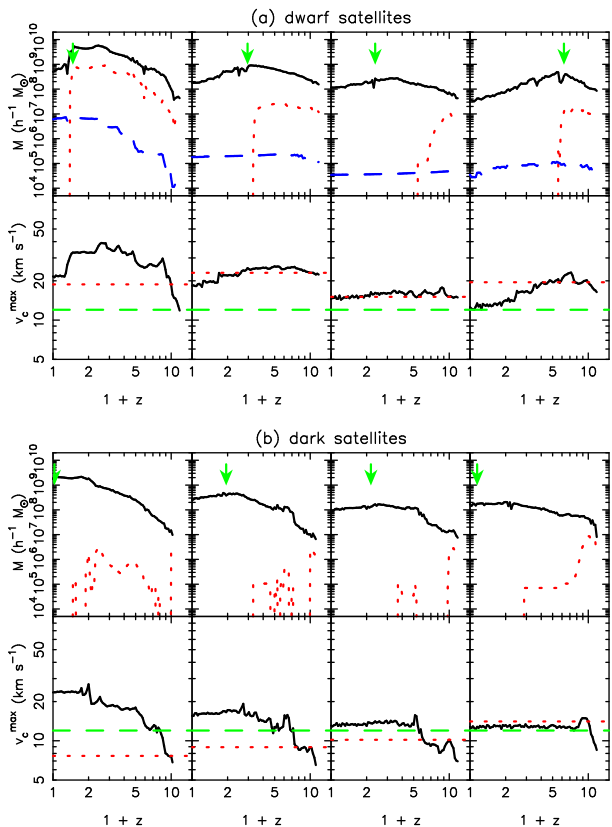}
\end{center}
\caption{ Formation histories of dwarf satellites and dark subhaloes. 
(a) total mass (solid), gas mass (dotted) and
stellar mass (dashed) in the main progenitors of a selection of
simulated haloes that end up as dwarf satellites today, plotted as a
function of $1 + z$. Reionisation occurs at $1 + z_{\rm re} =
10$. Arrows indicate the redshift when the progenitor is accreted into
a larger halo (i.e. when it becomes a subhalo). {\it Lower panels}:
the maximum of the circular velocity of the haloes. The horizontal
dotted line marks $v_{\rm c, 9}$ while the horizontal dashed line
indicates the critical velocity at the epoch of reionisation ($v_{\rm
c, 9} = 12$ km s$^{-1}$). (b) as (a), but for failed subhaloes that
end up as dark satellites today. }
\label{histories}
\end{figure}

In our simulations, reionisation is the main process that separates
successful from unsuccessful haloes. However, the properties of a
satellite are determined by subsequent events. Particularly important
is the accretion of the (sub)halo into the main progenitor of the host
galaxy. The epoch of accretion determines many of the key properties
of the final satellite, such as its luminosity and star formation
rate. The key events in the life of a subhalo are illustrated in
Fig.~\ref{histories}.

Consider first the evolution of haloes that go on to make visible
satellites by the present day, illustrated in panels~(a). As may be
seen in the lower subpanels, at $z_{\rm re} = 9$ all these haloes have
above-threshold maximum circular velocity ($v_{\rm c, 9} > 12$ km
s$^{-1}$). The gas in these objects, shown by the red line in the
upper subpanels, is unaffected by reionisation and these haloes are
able to form stars (blue lines). Eventually, however, the halo is
accreted into the progenitor halo of the main galaxy, becoming a
subhalo. At this time, the remaining gas is stripped completely and
star formation ceases. This explains why most of today's dwarf
satellites are old. Although we do not find the correlation between
accretion redshift and halo mass at the epoch of accretion reported in
the semi-analytic models of \citet{maccio09a}, we do find a
correlation between accretion redshift and final satellite
luminosity. The satellite illustrated in the 3rd column has a slightly
unusual behaviour. This object loses its gas well before becoming a
subhalo. Since its maximum circular velocity at $z=9$ is just above
the critical value, reionisation and photoheating do heat up the gas
which becomes loosely bound and is subsequently blown out by supernova
explosions and increased background radiation by $z \sim 6$. 
The photoionising background prevents gas from reaccreting into a halo 
with such small circular velocity \citep{ogt08}.  

Panel~(b) follows the fate of typical failed
satellites, i.e. of haloes that are below the circular velocity
threshold at reionisation and are destined to remain as dark subhaloes
today. Most of them lose all of their gas by evaporation at $z_{\rm
re}$. Although they can subsequently gain small amounts of gas as they
grow in mass (and circular velocity), the gas density in their shallow
potential wells is much too low for stars to form. The halo 
illustrated in the rightmost panel has value of $v_{\rm c, 9}$ which
is only just above threshold. As a result, it does not lose its
gas immediately after reionisation but holds on to it for some time
before finally losing it at $z \simeq 2$. This object never manages to
make stars because the large thermal pressure prevents gas from
attaining a sufficiently large density and so the (sub)halo remains
forever dark.

\section{Summary and discussion} 

In this Letter we have used a high resolution N-body hydrodynamic
simulation to investigate the formation histories of the satellites
that form in a Milky-Way type galaxies in a $\Lambda$CDM
universe. These simulations have been shown to reproduce the main
optical and chemical properties of the Local Group satellites,
including their luminosity function \citep{ofjt09}. In this study we
have focused on the properties that distinguish the small subset of
subhaloes that succeed in making a galaxy, from the vast majority
that remain dark. 

Our simulations naturally reproduce the observation that the Milky
Way satellites have very similar central densities irrespective of
their luminosity. They also show that there is a threshold circular
velocity today, $v_{\rm c, 0} \simeq 23$ km s$^{-1}$, above which
all subhaloes present today host satellites with the luminosities
typical of the classical satellites of the Milky Way ($L_{\rm V} \ge
2.6 \times 10^5 \ L_\odot$). The number of satellites drops rapidly
below this critical circular velocity but around this value there
are haloes which succeed in making a galaxy and haloes which do
not. The distinction between them can be traced back to the epoch of
reionisation when there is a hard threshold in maximum circular
velocity, $v_{\rm c, 9} \simeq 12$ km s$^{-1}$, that divides haloes
where gas can cool and make stars from haloes in which it
cannot. The scatter between $v_{\rm c, 0}$ and $v_{\rm c, 9}$ that
develops over time explains why seemingly similar subhaloes today
can have such different mass-to-light ratios. This key feature
cannot be explained by the effect of galactic winds alone. Galactic
winds in which the wind velocity scales in proportion to velocity
dispersion were shown by \citep{ofjt09} to be required to explain
the satellite luminosity function. Here we have shown that
reionisation is required to explain the circular velocity function.

Haloes with sub-threshold $v_{\rm c, 9}$ lose their gas by
evaporation during reionisation while the haloes with
above-threshold $v_{\rm c, 9}$ retain their gas which eventually
overcomes the increased thermal pressure produced by reionisation
and proceed to make stars. This explains why \citet{ofjt09} could
not find any residual signs of reionisation in the star formation
histories of simulated satellites.

If the CDM model is right, a large number of small, completely
optically dark subhaloes should be orbiting in the Milky Way halo. They
might one day be detected indirectly either through annihilation
radiation if the dark matter consists of supersymmetric particles
\cite[e.g.][]{Baltz00,Calc00,Evans04,Diemand07a,aquarius} or through 
gravitational lensing effects \cite[e.g][]{Kochanek91,Xu09}.

Broadly speaking, our results agree with those obtained using
semi-analytic models applied to high-resolution $N$-body
simulations. These studies also find that a preferred mass scale
arises naturally in dwarf galaxies from a combination of
astrophysical processes that inhibit star formation in small haloes
and the fact that the mass within 600 pc (or 300 pc) is not a very
sensitive function of the total subhalo mass \citep{li09,
maccio09a}.  It is nevertheless surprising that \citet{li09} managed
to obtain a correlation between $M_{06}$ and satellite luminosity
similar to ours (Fig.~\ref{strigari}) and to the real data because
they assumed an extreme reionisation model which is incompatible
with the results of recent simulations \citep{ogt08}. It is possible
that the minimum mass scale in their model was set by their
assumption that gas does not cool in haloes with virial temperature
below $10^4$K rather than by the effects of photoionising
background. Note that the threshold circular velocity imposed by
reionisation in our simulations, $v_{\rm c, 9} \simeq 12$ km
s$^{-1}$, corresponds to a virial temperature of $\sim 5 \times
10^3$K. 

Using a simple analytical model applied to an N-body simulation,
\cite{kravtsov09} also finds that a preferred satellite mass scale
(as well as other properties of the satellite system, such as the
luminosity function and spatial distribution) can be explained if
the star formation efficiency decreases with decreasing halo
mass. In our simulations a dependence of star formation efficiency
on halo mass of this kind is a natural outcome of the behaviour of
galactic winds. Kravtsov's model does not assume a sharp threshold
in the star formation efficiency. This, however, arises naturally in
our simulations as a result of reionisation and seems to be required
in order to explain why some dark matter halos always remain dark.

Our simulations ignore the possibilities that gas may be
self-shielded against ionising radiation, or that star formation may
result from molecular gas cooling before reionisation. Both of these
processes could be important in the formation of small objects
\citep{su04a, su04b} and may allow small amounts of star formation
in sub-threshold haloes. It will be particularly interesting to
explore how important these effects are for the newly-discovered
ultrafaint satellites \citep{kop08}. We leave these issues for
future work. Our present simulations, however, demonstrate that a
preferred mass scale in satellites does not require abandoning the
standard model of cosmology but is, in fact, a natural byproduct of
galaxy formation in small cold dark matter haloes.

\section*{Acknowledgments}

We thank Volker Springel for very useful comments on an early
draft. The simulations were carried out on the computational facilities at
the CCS, Tsukuba, the Cosmology Machine at the ICC, Durham, and Cray
XT4 at CfCA, NAOJ. This work was supported by the {\it FIRST}
project based on Grants-in-Aid for Specially Promoted Research by
MEXT (16002003), Grant-in-Aid for Scientific Research (S) by JSPS
(20224002), and an STFC rolling grant to the ICC. CSF acknowledges a
Royal Society Wolfson Research Merit Award.



\bsp

\label{lastpage}

\end{document}